\documentclass{cpbtex}

\usepackage{tikz}
\usepackage{tabularx}
\usepackage{listings}
\usepackage{cleveref}
\usepackage{makecell}

\makeatletter
\usepackage{jlcode}
\usepackage{amsmath}
\newtheorem{problem}{Problem}

\def\jlbasicfont{\ttfamily\footnotesize\selectfont}

\lstdefinestyle{code}{
  language=Julia, 
  showstringspaces=false,
  keywordstyle=\color{blue},
  commentstyle=\color{gray},
  identifierstyle=\color[RGB]{0,102,0},
  columns=fullflexible,
  keepspaces=true,
}
\lstnewenvironment{code}{\lstset{style=code}}{}
\DeclareMathAlphabet{\mymathbb}{U}{BOONDOX-ds}{m}{n}

\begin{document}

\title{Programming guide for solving constraint satisfaction problems with tensor networks}

\author{Xuanzhao Gao$^{1,2}$ and \ Xiaofeng Li$^{1}$ and \ Jinguo Liu$^{1}$\thanks{Corresponding author. E-mail:~jinguoliu@hkust-gz.edu.cn}\\
$^{1}${Hong Kong University of Science and Technology (Guangzhou), Guangzhou, China}\\
$^{2}${Hong Kong University of Science and Technology, Hong Kong SAR, China}
}   % The line break was forced via \\

\date{\today}
\maketitle

\begin{abstract}
    Constraint satisfaction problems (CSPs) are a class of problems that are ubiquitous in science and engineering. It features a collection of constraints specified over subsets of variables.
    A CSP can be solved either directly or by reducing it to other problems.
    This paper introduces the Julia ecosystem for solving and analyzing CSPs, focusing on the programming practices.
    We introduce some of the important CSPs and show how these problems are reduced to each other.
    We also show how to transform CSPs into tensor networks, how to optimize the tensor network contraction orders, and how to extract the solution space properties by contracting the tensor networks with generic element types. Examples are given, which include computing the entropy constant, analyzing the overlap gap property, and the reduction between CSPs.
\end{abstract}


\textbf{PACS: 02.10.Ox, 02.10.Xm, 01.50.hv} %no more than four PACS codes should be provided: check https://cpb.iphy.ac.cn/UserFiles/File/PACS2010Regular-Edition.pdf

% 02.10.Ox Combinatorics; graph theory
% 02.10.Xm Multilinear algebra
% 01.50.hv Computer software and software reviews

\section{Introduction}\label{sec:intro}

A constraint satisfaction problem (CSP) is a class of problems that are ubiquitous in science and engineering.
These problems include, for example, the independent set problem~\cite{Clark1991}, the cutting problem~\cite{ding2001min}, dominating set~\cite{Clark1991}, set packing~\cite{crescenzi1995compendium}, set covering~\cite{chvatal1979greedy}, vertex coloring~\cite{jensen2011graph, malaguti2010survey}, K-SAT~\cite{biere2009handbook,schaefer1978complexity},  and the vertex cover problem~\cite{Moore2011}.
These problems have a wide range of applications in scheduling, logistics, wireless networks and telecommunication, and computer vision, among others~\cite{Butenko2003, Wu2015}.
Finding an optimum solution for these problems is typically NP-hard in the worst case~\cite{Hastad1996}.

How to solve these problems efficiently is a longstanding challenge in computer science, and is more and more connected to physics.
The performance of an algorithm is closely related to the problem size and the solution space properties or energy landscape~\cite{Mezard1984}.
For example, if a problem exhibits the overlap gap property, then any local search algorithm may fail to find the global optimum in sub-exponential time~\cite{Gamarnik2021}.
It also applies to quantum algorithms, for example, the quantum variational optimizer for the independent set problem can be polynomially faster than the classical algorithm when the solution space is uniformly connected~\cite{Cain2023, Ebadi2022}.
So understanding the solution space properties is crucial for designing efficient classical and quantum algorithms.
A CSP can also be addressed by reducing it to other CSPs so that solvers for these problems can also be used.
In this case, the overhead of the reduction is critical for the performance as well.
Efficient reduction algorithms are developed in recent years due to the rising trend in using physical systems to solve general CSPs through reductions to other problems, such as spin glass on Ising machines~\cite{Mohseni2022, lucas2014ising}, QUBO on quantum annealing processor~\cite{Ushijima-Mwesigwa2017} and independent set problems on King's sub-graph~\cite{Pichler2018,Ebadi2022,Nguyen2023}.
While problem reductions are fundamental concepts in computer science, these techniques are not widely used by physicists in the past.

\begin{figure}[h]
    \centering
    \includegraphics[width=0.6\textwidth, trim={0cm 0cm 0cm 0cm}, clip]{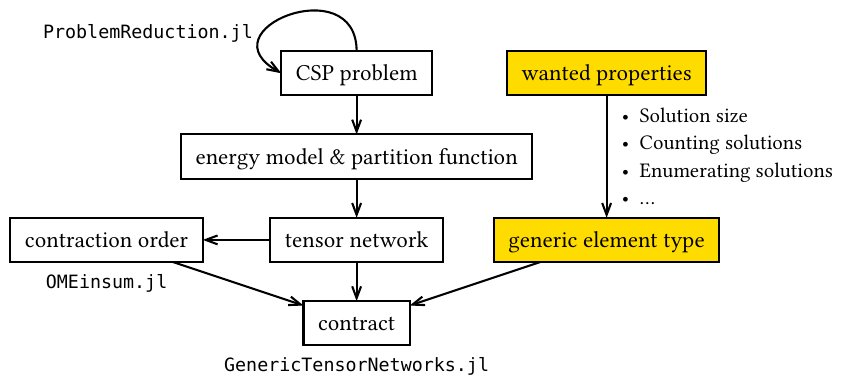}
    \caption{The generic tensor network framework for solution space analysis of constraint satisfaction problems (CSPs).}
    \label{fig:gtn}
\end{figure}

To reduce the barrier for physicists to study CSPs, we introduce the Julia~\cite{Bezanson2012} ecosystem for reducing and analyzing CSPs.
Julia is a high performance programming language designed scientific computing. 
It is as easy to use as Python, but with the performance of C~\cite{Jeff2015}.
The main tool we will use is \texttt{GenericTensorNetworks.jl}, which is a tensor network based CSPs solution framework~\cite{Liu2021,Liu2023}. 
At the time of writing, its version is 3.2.0.
Numerous other scientific computing software packages have been developed based on Julia, such as the {ITensors.jl}~\cite{Fishman2022} for tensor network based many-body physics, the \texttt{Yao.jl}~\cite{Luo2020} for variational quantum circuits, \texttt{JuMP.jl}~\cite{Lubin2023} for mathematical optimization, and \texttt{DifferentialEquations.jl}~\cite{Rackauckas2017} for solving differential equations.
The \href{https://github.com/QuEraComputing/GenericTensorNetworks.jl}{\texttt{GenericTensorNetworks.jl}} enables us to extract the solution space properties by contracting the tensor networks with generic element types, where the supported solution space properties include the partition function, the solution size, the number of solutions, the solution enumeration and sampling.
The framework is shown in \Cref{fig:gtn}.
\texttt{GenericTensorNetworks.jl} is built on top of the tensor network contraction library \href{https://github.com/under-Peter/OMEinsum.jl}{\texttt{OMEinsum.jl}} and the problem reduction library \href{https://github.com/GiggleLiu/ProblemReductions.jl}{\texttt{ProblemReductions.jl}}.
The \texttt{ProblemReductions.jl} library provides a set of problem definitions and reduction interfaces for CSPs.
The \texttt{OMEinsum.jl} library provides the tensor network contraction engine, with the state-of-the-art contraction order optimization technique.
Its tensor network contraction supports the CUDA backend, which is powered by Julia's dynamic and generic approach for GPU programming~\cite{Besard2018}.
% The \texttt{GenericTensorNetworks.jl} library converts the CSPs into tensor networks, optimize the contraction order, and extract the solution space properties by contracting the tensor networks with generic element types.

In the following discussion, we first introduce the CSP problems in \Cref{sec:csp}\hspace{1.2em}, with a focus on the reduction between CSPs.
Then we introduce the tensor network representation of CSPs in \Cref{sec:tensor}\hspace{1.2em}, and show how to optimze the tensor network contraction order in \Cref{sec:contraction}\hspace{1.2em}.
Finally, we show how to extract the solution space properties by contracting the tensor networks with generic element types in \Cref{sec:solution}\hspace{1.2em}.
If you have a Julia REPL, you can follow the code examples in the following sections through copy-pasting the codes.

\section{Constraint satisfaction problem}\label{sec:csp}

\begin{figure}[h]
    \centering
    \includegraphics[width=0.7\textwidth, trim={0 0 0 0}, clip]{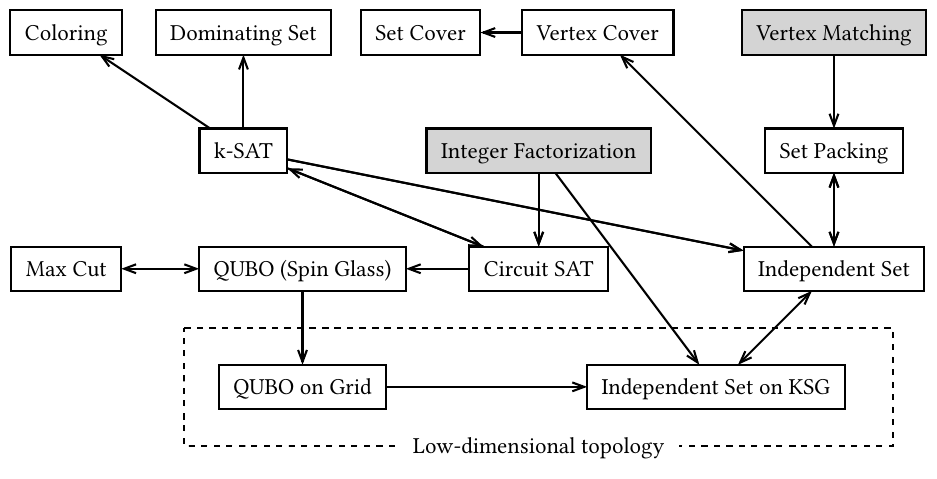}
    \caption{A gallery of CSPs and their reductions in \texttt{ProblemReductions.jl}. The arrows denote the reductions. The problem pointed by the tail can be reduced to the problem pointed by the head.
    Only the problems with gray background are not in NP-complete.
    }
    \label{fig:reduction}
\end{figure}

% The pursuit of solving a computational hard problem more efficiently is persistent.
Among the computational hard problems, the constraint satisfaction problems (CSPs) are a class of problems that are closely related to physical models.
\texttt{ProblemReductions.jl} provides an interface for defining CSPs and reductions between them as shown in \Cref{fig:reduction}.
In this section, we will introduce the definition of some famous CSPs, and show how these problems are defined and reduced to each other in \texttt{ProblemReductions.jl}.

\subsection{Problems definitions}
The CSPs problems implemented in \texttt{ProblemReductions.jl} are listed in \Cref{fig:reduction}, below we will introduce these problems one by one.

\subsubsection{Problems on graphs}

\begin{figure}[h]
    \centering
    \includegraphics[width=0.9\textwidth, trim={0cm 0cm 0cm 0cm}, clip]{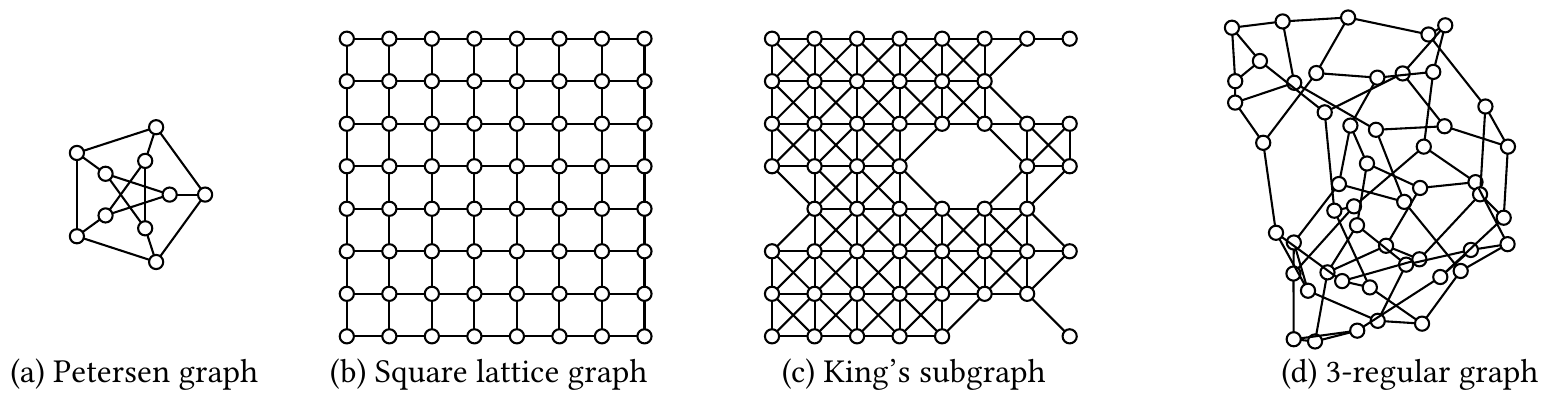}
    \caption{The types of graphs used in the partition function calculation. (a) is the famous Petersen graph, (b) is a square lattice graph, (c) is a King's sub-graph, and (d) is a random 3-regular graph, where each vertex has degree 3.}
    \label{fig:graphtypes}
\end{figure}

We start by defining the graph topologies used in this work.
Most CSPs introduced in this work are defined on graphs, which are not limited to simple graphs.
A simple graph can be relaxed to \emph{hyper-graphs}, where an edge can connect any number of vertices, or restricted to \emph{unit disk graphs}, where an edge can only connect two vertices with distance less than a given threshold in a low dimensional embedding space.
To simplify the discussion, we will focus on unit disk graphs in \Cref{fig:graphtypes} (b) and (c), and simple graphs in \Cref{fig:graphtypes} (a) and (d).
Their code implementation is as follows.
\begin{jllisting}
julia> using Graphs, GenericTensorNetworks
julia> petersen = Graphs.smallgraph(:petersen);
julia> square(L) = GenericTensorNetworks.random_square_lattice_graph(L, L, 1.0);
julia> ksg(L) = GenericTensorNetworks.random_diagonal_coupled_graph(L, L, 0.8);
julia> regular3(n) = Graphs.random_regular_graph(n, 3);
\end{jllisting}

% First, we will introduce the problems that are defined on graphs.

\textbf{Spin glass}~\cite{Glover2018,Qiu2020} system is a type of disordered magnetic system that exhibits glassy behavior.
The Hamiltonian of the system on a simple graph $G=(V, E)$ is given by
\begin{equation}
H(G, \sigma) = \sum_{(i,j) \in E} J_{ij} \sigma_i \sigma_j + \sum_{i \in V} h_i \sigma_i
\end{equation}
where $J_{ij} \in \mathbb{R}$ is the coupling strength between spins $i$ and $j$, $h_i \in \mathbb{R}$ is the external field on spin $i$, and $\sigma_i \in \{-1, 1\}$ is the spin variable.
The following code defines a spin glass on a Petersen graph with unit coupling strength and zero external field.
\begin{jllisting}
julia> using ProblemReductions
julia> spin_glass = SpinGlass(
           petersen,     # graph
           ones(15),     # J, in order of edges
           zeros(10)     # h, in order of vertices
       );
julia> energy(spin_glass, [-1,-1,-1,-1,-1,1,1,1,1,1])
5.0
\end{jllisting}
One can get detailed description from the docstring of the function. Just type \texttt{?} in the Julia REPL followed by the function name, e.g. What's more, we could use \texttt{fieldnames} function to print the fields of a problem type.
\begin{jllisting}
julia> ?SpinGlass
julia> fieldnames(SpinGlass)
(:graph, :J, :h)
\end{jllisting}

A equivalent problem to the spin glass problem is the \textbf{Quadratic unconstrained binary optimization (QUBO)}~\cite{Glover2018}, which can be described as a quadratic form $f_Q(x) = x^{T} Q x$, where $x$ is a vector of binary decision variables and $Q$ is a square matrix. 
The matrix $Q$ can be interpreted as the adjacency matrix of a graph, where the diagonal elements are zero.
Finding the assignment for each entry of $x$ such that $f_Q(x)$ is minimized (maximized) is a NP-hard problem and cannot be solved efficiently by any existing algorithm.
Due to its close connection to Ising models, QUBO constitutes a central problem class for adiabatic quantum computation, where it is solved through a physical process called quantum annealing.
In \texttt{ProblemReductions.jl}, we could use type \texttt{QUBO} to define the QUBO problem.
\begin{jllisting}
julia> Q = [1 1; 0 0]
2×2 Matrix{Int64}:
 1  1
 0  0

julia> QUBO(Q);  # matrix Q as argument
\end{jllisting}

Another related NP-hard problem is the \textbf{Max cut}~\cite{ding2001min}, which is to partition the vertices into two subsets that maximizes the total weight of the edges connecting the two subsets.
This problem is equivalent to the \textbf{Spin glass} problem with positive uniform coupling strength and zero external field, and the ground state of the spin glass problem is the solution to the max cut problem.
An example is given below, where a max cut problem is defined on a Petersen graph.
\begin{jllisting}
julia> MaxCut(
           petersen,                # graph
           UnitWeight(ne(petersen)) # weights of edges
       );
\end{jllisting}

%The hardness of a problem is closely related to the domain of variables and the graph topology.
% This definition naturally extends to the case of a hyper-graph:
% \begin{equation}
% H(G, \sigma) = \sum_{e \in E} J_{e} \prod_k \sigma_k + \sum_{i \in V} h_i \sigma_i,
% \end{equation}
% where $J_e$ is the coupling strength associated with hyperedge $e$, and the product is over all spins in the hyperedge.
\textbf{Independent sets}~\cite{Clark1991} are the subsets of the vertices of the graph, where no two of which are adjacnet, and the independent set problem is to find such sets of a given graph.
The problem is closely related to the hard-core lattice gas model~\cite{Scott2005} in statistical physics.
A \textbf{maximum independent set (MIS)} is an independent set of largest possible size for a given graph $G$, and the optimization problem of finding such a set is called the maximum independent set problem.
Finding an MIS for a given graph is a NP-complete problem~\cite{tarjan1977finding} and various exponential-time algorithms have been developed for solving it~\cite{xiao2013}.
It has garnered significant attention in recent decades due to its natural mapping to Rydberg atom-based quantum computing~\cite{Nguyen2023}.
The following code defines the independent set problem on a Petersen graph.
\begin{jllisting}
julia> IndependentSet(
           petersen,                # graph
           UnitWeight(nv(petersen)) # weights of vertices
       );
\end{jllisting}
The second field is the weights of the vertices, which are set to \texttt{UnitWeight} by default, representing the unweighted version.
The \textit{size} of the independent set is the sum of the weights of its vertices.
In the following, we will not explicitly state this for each problem in the rest of the paper for the sake of simplicity.
A variant of the independent set problem is the \textbf{maximal independent set (MIS)} problem, where the ``maximal'' means that the independent set can not be extended by adding any other vertex.
Its solution space contains all the maximal independent sets, which is smaller than the solution space of the independent set problem.
\begin{jllisting}
julia> MaximalIS(
           petersen,                # graph
           UnitWeight(nv(petersen)) # weights of vertices
       );
\end{jllisting}
Another problem that is closely related to the independent set problem is the \textbf{minimum vertex cover} problem.
They are treated as the same problem in many literatures.
A vertex cover is a set of vertices such that every edge in the graph has at least one endpoint in this set, and the minimum vertex cover problem is to find the smallest such set.
The complement of an independent set must be a vertex cover, since an edge can not be covered by two vertices in the independent set.
Conversely, the complement of a vertex cover is an independent set.
Therefore, the complement of the maximum independent set is the minimum vertex cover.
In the following example, we define the minimum vertex cover problem on a Petersen graph.
\begin{jllisting}
julia> VertexCovering(
            petersen,                # graph
            UnitWeight(nv(petersen)) # weights of vertices
        );
\end{jllisting}

\textbf{Vertex coloring}~\cite{jensen2011graph, malaguti2010survey} refers to the problem of assigning colors to the vertices of a graph in such a way that no two adjacent vertices share the same color. The vertex coloring problem seeks to determine whether it is possible to color all vertices of a given simple graph using $k$ colors while ensuring that adjacent vertices are not colored the same. 
The problem to find the minimum number of colors required to color a given graph is one of Karp's 21 NP-complete problems~\cite{karp2010reducibility}.
One of the major applications of graph coloring is register allocation in compiler~\cite{Chaitin1982}.
In the following example, we define the vertex coloring problem on a Petersen graph using 3 colors.
\begin{jllisting}
julia> Coloring{3}(                 # 3 kinds of colors
           petersen,                # graph
           UnitWeight(ne(petersen)) # weights on edges
       );
\end{jllisting}

\textbf{Dominating set}~\cite{Clark1991} is a problem to identify a dominating set that minimizes either the number of vertices or the total weight. A dominating set for a graph $G$ is a subset $D$ of its vertices such that every vertex in $G$ is either included in $D$ or is adjacent to at least one vertex in $D$. 
It is a classical NP-complete decision problem and has a wide application in fields such as wireless networking~\cite{Stojmenovic2002}, document summarization, and the designing secure systems for electrical grids.
Below is the code that defines the dominating set problem on a Petersen graph.
\begin{jllisting}
julia> DominatingSet(
           petersen,                # graph
           UnitWeight(nv(petersen)) # weights of vertices
       );
\end{jllisting}

\textbf{Matching}~\cite{Lovasz2009} is a subset of the edges of the graph such that no two edges in the set have the same vertex, and the vertex matching problem is to find the matching with maximum number of edges or maximum total weight.
Finding the maximum matching for a simple graph $G = (V, E)$ can be solved by Edmonds' blossom algorithm in $O(|V|^2|E|)$ time~\cite{Edmonds1965}, hence no hard problem can be reduced to the matching problem. 
However, the counting version of the matching problem is in the complexity class \#P-complete~\cite{Valiant1979}.
The following code defines the matching problem on a Petersen graph.
\begin{jllisting}
julia> Matching(
           petersen,                # graph
           UnitWeight(ne(petersen)) # weights of edges
       );
\end{jllisting}

\subsubsection{Boolean satisfiability problems}

\textbf{Satisfiability problem}~\cite{biere2009handbook,schaefer1978complexity} is the first problem that was proven to be NP-complete, which is to determine whether a given boolean formula is satisfiable.
The satisfiability problem is a fundamental problem in computer science and has wide applications in fields such as artificial intelligence, cryptography, and automated reasoning.
There is no known algorithm can solve SAT problem in time faster than exponential-time~\cite{baumgartner2002first, biere2009conflict}.
In \texttt{ProblemReductions.jl}, the boolean variables, i.e. the literals, are defined by the \texttt{BoolVar} function as follows.
\begin{jllisting}
julia> x, y, z, a, b = BoolVar.(["x", "y", "z", "a", "b"]); # variables
\end{jllisting}
% The fields for Satisfiability are: \texttt{symbols},\texttt{weights},\texttt{cnf}.
% The \texttt{cnf} is the conjunctive normal form of the boolean formula, the \texttt{variables} is the list of variables in the formula, and the \texttt{weights} is defined on each clause and is \texttt{UnitWeight} by default.
With these boolean variables, a general satisfiability problem is defined as follows, where the boolean formula is given in the conjunctive normal form (CNF).
\begin{jllisting}
julia> cnf = (x ∨ y ∨ z) ∧ (a ∨ b ∨ ¬y ∨ z) ∧ (¬x ∨ ¬a)
(x ∨ y ∨ z) ∧ (a ∨ b ∨ ¬y ∨ z) ∧ (¬x ∨ ¬a)
julia> ?∨     # type `?` to enter help mode
"∨" can be typed by \vee<tab>
julia> sat_problem = Satisfiability(
           cnf,                    # CNF
           UnitWeight(length(cnf)) # weights of clauses
       );
\end{jllisting}
If the number of literals in a clause is fixed to $k$, using the \textbf{K-satisfiability problem} is preferred.
In the following example, we define a 3-SAT problem.
\begin{jllisting}
julia> ksat_problem = KSatisfiability{3}((x ∨ y ∨ z) ∧ (a ∨ b ∨ ¬y) ∧ (¬x ∨ ¬a ∨ ¬b));
\end{jllisting}
\textbf{Circuit satisfiability (Circuit SAT)}~\cite{Moore2011} is a variant of the satisfiability problem, where the boolean formula is represented by a circuit.
% The fields for CircuitSAT are: \texttt{circuit},\texttt{symbol}, \texttt{weights}. Weights are defined on each boolean expression and are \texttt{UnitWeight} by default. 
In \texttt{ProblemReductions.jl}, we could use macro expression \texttt{@circuit} to define a circuit and then use it as the input of the CircuitSAT problem.
\begin{jllisting}
julia> circuit = @circuit  begin
           c = x ∧ y
           d = x ∨ (c ∧ ¬z)
           d = true
       end
Circuit:
| c = ∧(x, y)
| d = ∨(x, ∧(c, ¬(z)))
| d = true
julia> CircuitSAT(circuit);
\end{jllisting}

\subsubsection{Other problems}

\textbf{Integer factorization (factoring) problem}~\cite{Moore2011, montgomery1994survey} is to find the decomposition of a positive integer into product of integers. 
The corresponding decision problem is to determine whether a given integer $n$ has a factor, other than $1$, that is smaller than $k$, which is classified as NP-intermediate (less difficult than NP-complete but more difficult than P).
Many cryptographic protocols are based on the presumed difficulty of factoring large composite integers or a related problem—for example, the RSA problem~\cite{mumtaz2019forty}.
An algorithm that efficiently factors an arbitrary integer would render RSA-based public-key cryptography insecure. 
In \texttt{ProblemReductions.jl}, one can define the factoring problem by the \texttt{Factoring} function as follows.
\begin{jllisting}
julia> Factoring(
           2,  # the number of bits to store the first factor
           3,  # the number of bits to store the second factor
           15  # the integer to be factored
       );
\end{jllisting}

\textbf{Set cover problem}~\cite{chvatal1979greedy} is defined on a set of elements and a collection of subsets, and the goal is to find the minimum number of subsets that cover all the elements.
It is a hypergraph generalization of vertex cover problem, however, its decision version is in the complexity class NP-complete~\cite{karp2010reducibility}.
An example is given below, where the set of elements is $\left\{1,2,3,4,5\right\}$ and the subsets are $\left\{\{1,2,3\},\{1,3,5\},\{1,2\},\{4\},\{4,5\}\right\}$.
\begin{jllisting}
julia> subsets = [[1,2,3],[1,3,5],[1,2],[4],[4,5]];
julia> SetCovering(
           subsets,                    # a collection of subsets
           UnitWeight(length(subsets)) # weights of subsets
       );
\end{jllisting}

\textbf{Set packing}~\cite{crescenzi1995compendium} is also defined on a set of elements and a collection of subsets. The goal is to find the maximum number of subsets that are pairwise disjoint, meaning that no two subsets share any common elements. It is a hypergraph generalization of the independent set problem, and its decision version is also in NP-complete~\cite{karp2010reducibility}.
The following example shows how to define a set packing problem.
\begin{jllisting}
julia> sets = [[1,2,3],[1,3,5],[1,2],[4],[4,5]];
julia> SetPacking(
           sets,                    # a collection of sets
           UnitWeight(length(sets)) # weights of sets
       );
\end{jllisting}

% \textbf{(Binary) Paint shop problem} is an optimization problem. An intuitive illustration for the problem is given a length $2m$ sequence containing $m$ cars, each appearing twice, we need to color each cars with 2 colors. Our goal is to determine when to color the car with which kind of colors so that we minimize the times of changing the current color. The fields for PaintShop are: \texttt{sequence},\texttt{isfirst}. The field \texttt{isfirst} is an extra information indicating whether the symbol in the sequence appears for the first time and it is generated by the constructor. A simple example is \texttt{sequence = ["a","b","a","c","c","b"]}, the corresponding \texttt{isfirst = [1,1,0,1,0,0]}
% \begin{jllisting}
% julia> PaintShop(["a","b","a","c","c","b"])
% PaintShop{String}(["a", "b", "a", "c", "c", "b"], Bool[1, 1, 0, 1, 0, 0])
% \end{jllisting}

\subsection{Reduction interfaces}

In this section, we introduce the interfaces for problem reduction in \texttt{ProblemReductions.jl}.
If a problem $A$ can be reduced to problem $B$, then we can use the solution of $B$ to solve $A$, i.e. $A \leq_p B$, meaning $B$ is not easier than $A$.
The reduction is not only a tool to show the hardness of a problem, but also a tool to connect different problems. For example, it allows us to solve a problem by reducing it to an Ising problem or an independent set problem on low dimensional grid that implementable on physical hardware, e.g. D-Wave quantum annealing processor, Ising machines~\cite{Mohseni2022} and Rydberg atom arrays~\cite{Pichler2018,Ebadi2022,Nguyen2023}.

In \Cref{fig:reduction}, we show some common reductions between constraint satisfaction problems (CSPs).
Since it is not practical to construct the reduction between all the CSPs, in \texttt{ProblemReductions.jl}, we only implement a finite set of reductions.
A problem can be reduced to another if a reduction path exists to connect them.

\begin{figure}[h]
    \centering
    \includegraphics[width=0.8\textwidth, trim={0cm 0cm 0cm 0cm}, clip]{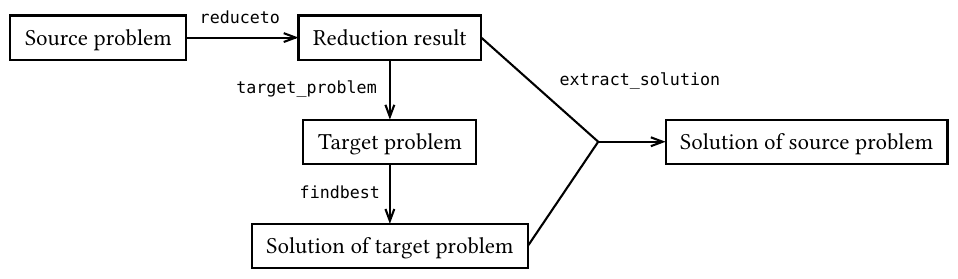}
    \caption{The workflow of problem reduction in \texttt{ProblemReductions.jl}.}
    \label{fig:reduction_workflow}
\end{figure}

As shown in \Cref{fig:reduction_workflow}, a workflow of problem reduction can be separated into four parts: reduction, target problem extraction, solving, and solution extraction.
In the reduction phase, we call the \texttt{reduceto} function to reduce a source problem to a target problem.
The reduction result contains not only the target problem, but also intermediate information to help convert the solution back.
The target problem can be extracted by \texttt{target\_problem} function. 
To solve it, we resort to the unified solver interface \texttt{findbest}.
Finally, we could extract the solution for source problem by \texttt{extract\_solution} function.
In the following example, we convert an integer factorization problem to a circuit satisfiability problem with a direct reduction rule.
\begin{jllisting}
julia> using ProblemReductions, GenericTensorNetworks
julia> factoring = Factoring(2, 3, 15);
julia> reduction_res = reduceto(CircuitSAT, factoring);  # direct reduction
julia> result = findbest(
           target_problem(reduction_res),    # target problem
           GTNSolver()                       # solver
       ); 
julia> solution = extract_solution(reduction_res, result[1]);
julia> ProblemReductions.read_solution(factoring,solution) # returns two factors
(3, 5)
\end{jllisting}
Here, the \texttt{findbest} function is used to solve the circuit satisfiability problem.
The built-in solver is \texttt{BruteForce}. After \texttt{GenericTensorNetworks.jl} is loaded, a tensor network based solver \texttt{GTNSolver} will also be available.

In some other cases, a direct reduction rule is not available, then a reduction path must be provided by the user.
It can be constructed by the \texttt{reduction\_paths} function that returns all the possible reduction paths connecting two problem types.
In the following example, we convert an integer factorization problem to a spin glass problem with an automatically generated reduction path.
\begin{jllisting}
julia> all_paths = reduction_paths(Factoring, SpinGlass)
julia> reduction_res = reduceto(first(all_paths), factoring)
julia> spin_glass = target_problem(reduction_res);
julia> problem_size(spin_glass)
(num_vertices = 63, num_edges = 137)
julia> result = findbest(spin_glass, GTNSolver()); # returns all best solutions
julia> solution = extract_solution(reduction_res, result[1]);
julia> ProblemReductions.read_solution(factoring,solution) # returns two factors
(3, 5)
\end{jllisting}
In this example, there exists multiple reduction paths from the factoring problem to the spin glass problem, which is \texttt{Factoring} $\rightarrow$ \texttt{CircuitSAT} $\rightarrow$ \texttt{SpinGlass}~\cite{Nguyen2023}. 
The reduction transforms the factoring problem into $63$ vertices spin glass problems.
Although the reduction is not free and requires some overhead, it is a powerful tool that enables us to fully utilize existing solvers to solve new problems.

\section{Tensor network representation of constraint satisfaction problems}\label{sec:tensor}

In previous sections, we have introduced the CSPs and the reductions between them, but a general framework to solve these problems is still missing.
In this section, we introduce a tensor network representation of the constraint satisfaction problems (CSPs) to facilitate their analysis and solution.
We show that both energy model and partition function of the CSPs can be effectively represented by tensor networks, and an example tensor network representation of the independent set problem is given.

\subsection{Energy model and partition function}

Let us consider a constraint satisfaction problem on a hyper-graph $G=(V, E)$, where $V$ is the set of vertices that associated with variables and $E$ is the set of hyper-edges that associated with constraints.
The energy model of the problem is defined as
\begin{equation}\label{eq:eng}
    \mathcal{E}(G, \mathbf{s}) = \sum_{e \in E} h(e, \mathbf{s})
\end{equation}
where $\mathbf{s}$ is an assignment of variables and $h(e, \mathbf{s})$ is an energy term associated with hyper-edge $e \in E$ representing the constraint on $e$.
Invalid configurations due to the hard constraints in some problems, e.g. the independence constraint in the independent set problem, are characterized by infinite energy penalty.
The partition function for the energy model at inverse temperature $\beta$ is defined as
\begin{equation}\label{eq:partition}
    Z(G, \beta) = \sum_{\mathbf{s}}e^{-\beta \mathcal{E}(G, \mathbf{s})} = \sum_{\mathbf{s}} \prod_{e \in E} e^{ - \beta h(e, \mathbf{s})}\;.
\end{equation}
Since the energy term $h(e, \mathbf{s})$ only involves the spins on the vfrtices that belong to the hyper-edge $e$,
$e^{\beta h(e, \mathbf{s})}$ is a tensor of rank $|e|$ and the partition function is a tensor network, where two tensors are connected if and only if they have shared variables in their corresponding hyper-edges.
In the infinite temperature limit, $\beta \rightarrow 0$, the partition function is equivalent to the number of valid configurations.
For certain problems, although finding one best solution is easy, counting the number of valid configurations is hard,
e.g. counting the number of satisfying assignments of 2-SAT is \#P-hard~\cite{Valiant1979}.

\subsection{Example: Tensor network representation of the independent set problem}\label{sec:independent_set}

In the following, we will show the tensor network representation of the independent set problem.
Given a graph $G = (V, E)$, the energy model of the independent set problem is:
\begin{equation}\label{eq:independent_set}
    H(G, \mathbf{n}) = - \sum_{v \in V} w_v n_v + \sum_{(u, v) \in E} \infty n_u n_v
\end{equation}
where $n_v$ is the number of vertices in the independent set, i.e. $n_v = 1$ if $v$ is in the independent set, and $n_v = 0$ otherwise, $w_v$ is the weight of vertex $v$.
The larger the size of the independent set, the lower the energy. An example is given below. The partition function at inverse temperature $\beta$ can be represented as
\begin{equation}
    Z(G, \beta) = \sum_{\mathbf{n}} \prod_{(u, v) \in E} \underbrace{e^{ - \beta \infty n_u n_v}}_{B(n_u, n_v)}\prod_{v \in V} \underbrace{e^{\beta w_v n_v}}_{W(n_v)}\;.
\end{equation}
It can be represented as a tensor network.
% Let us denote $x = e^{\beta}$, 
For each edge $(u, v) \in E$, the pairwise interaction can be represented as a rank 2 tensor
\begin{equation}
    B(n_u, n_v) = \begin{pmatrix}
        \mymathbb{1} & \mymathbb{1}\\
        \mymathbb{1} & \mymathbb{0}
    \end{pmatrix}\;,
\end{equation}
where $\mymathbb{1}$ and $\mymathbb{0}$ are the multiplicative identity and zero, respectively.
For each vertex $v \in V$, a rank-one tensor $W(n_v)$ is associated with each vertex $v \in V$, which is given by
\begin{equation}
    W(n_v) = \begin{pmatrix}
        \mymathbb{1} \\
        e^{\beta w_v}
    \end{pmatrix}\;.
\end{equation}
As an example, we consider computing the partition function of a square graph with 4 vertices and 4 edges.
The graph and its corresponding tensor network for the independent set problem are shown bellow.\newline
\centerline{\includegraphics[width=0.4\textwidth]{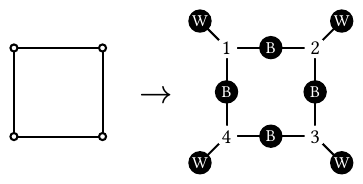}}
The resulting tensor network is equivalent to the following sum-product form
\begin{equation}
    Z = \sum_{n_1, n_2, n_3, n_4} B(n_1, n_2) B(n_1, n_4) B(n_2, n_3) B(n_3, n_4) W(n_1) W(n_2) W(n_3) W(n_4)\;.
\end{equation}
Using \texttt{GenericTensorNetworks}, the tensor network can be constructed automatically as follows.
\begin{lstlisting}
julia> using GenericTensorNetworks, OMEinsum, Graphs
julia> problem = IndependentSet(square(2), ones(4));
julia> energy(problem, [1, 1, 1, 0])  # violation of hard constraints
Inf
julia> energy(problem, [1, 0, 0, 1])
-2.0
julia> network = GenericTensorNetwork(problem)
GenericTensorNetwork{IndependentSet{SimpleGraph{Int64}, Int64, UnitWeight}, DynamicNestedEinsum{Int64}, Int64}
- open vertices: Int64[]
- fixed vertices: Dict{Int64, Int64}()
- contraction time = 2^5.17, space = 2^2.0, read-write = 2^6.19
julia> fieldnames(typeof(network))
(:problem, :code, :fixedvertices)
julia> OMEinsum.flatten(network.code)
1∘2, 1∘4, 2∘3, 3∘4, 1, 2, 3, 4 ->
\end{lstlisting}
The resulting tensor network has 3 fields: \texttt{problem}, \texttt{code}, and \texttt{fixedvertices}.
The \texttt{problem} field is the problem instance of type \texttt{IndependentSet}.
The \texttt{code} field is the einsum notation defined by \texttt{OMEinsum} with optimized contraction order.
Einsum notation is a compact way to represent tensor networks topology, and the contraction order is optimized by \texttt{OMEinsum} to minimize the number of operations and memory usage.
The \texttt{OMEinsum.flatten} function is used to remove the contraction order, such that we can see the underlying tensor network structure more clearly.
In the flattened form, the input tensors and output tensor are separated by the arrow symbol ``\texttt{->}''.
The output tensor is associated with an empty string, which means the output tensor is a scalar.
The input tensors are separated by commas, and the indices of the input tensors are separated by the ``\texttt{$\circ$}'' symbol.
In this example, there are 4 input tensors of rank 2 and 4 input tensors of rank 1.
The rank 2 tensors are associated with the edge tensors $B(n_u, n_v)$, and the rank 1 tensors are associated with the vertex tensors $W(n_v)$.
The last field \texttt{fixedvertices} is a dictionary that stores the fixed vertices and their corresponding values.
In this example, there are no fixed vertices, so the dictionary is empty.

\section{Choose the right tensor network contraction order optimizer}\label{sec:contraction}

After representing CSPs as tensor networks, we can extract the solution space properties by contracting the tensor networks with generic element types.
However, contracting a tensor network can be a challenging task, naively looping over all indices is $O(2^n)$ in time complexity, where $n$ is the number of indices.
To reduce the complexity, we need to find a good contraction order, which is the order of contracting the indices.
Different contraction orders lead to different complexities, while finding the optimal contraction order, i.e., the contraction order with minimal complexity, is NP-complete~\cite{Markov2008}.
Luckily, a close-to-optimal contraction order is usually acceptable, which could be found in a reasonable time with a heuristic optimizer.
In the past decade, methods have been developed to optimize the contraction orders, including both exact ones and heuristic ones.
Among these methods, multiple heuristic methods can handle networks with more than $10^4$ tensors efficiently~\cite{Gray2021, martin2024probabilistic}.
In this section, we will first introduce the basic concepts and then review the methods for finding a good contraction order.

\subsection{Contraction order optimization}

A contraction order can be represented as a rooted tree, where the leaves are the tensors to be contracted and the root is the final result.
In actual calculation, we prefer binary contractions, i.e., contracting two tensors at a time, so that we can make use of BLAS~\cite{lawson1979basic} libraries to speed up the calculation by converting these two tensors as matrices.
In this way, a given contraction order can be represented as a binary tree,
where the leaves are the original tensors, the nodes are the intermediate tensors after contracting two tensors, and the root is the final result.
For a given contraction order, the following three metrics are defined to describe its quality:
\begin{itemize}
    \item Time complexity (\texttt{tc}): total number of operations.
    \item Space complexity (\texttt{sc}): the maximum number of elements in the largest intermediate result.
    \item Read-write complexity (\texttt{rwc}): total number of elements to be read and written from memory.
\end{itemize}
The goal of contraction order optimization is to find a binary contraction order, with minimal time complexity or space complexity, which is called the \textbf{optimal contraction order}.
In the example shown in \Cref{sec:independent_set}\hspace{1.2em}, the optimal contraction order is given by the following binary tree.\newline
\centerline{\includegraphics[width=0.5\textwidth]{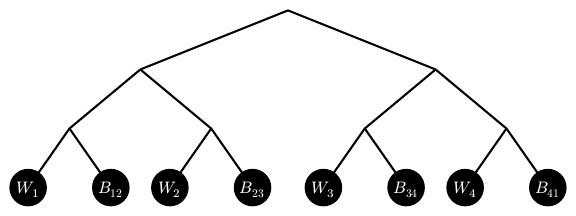}}
The largest tensor during contraction has a rank of $2$, which is the minimum among all possible contraction orders.
In the following, we define the unoptimized tensor network topology manually with \texttt{OMEinsum} and then show how to optimize the contraction order.

\begin{jllisting}
julia> using OMEinsum
julia> eincode = EinCode([[1, 2], [2, 3], [3, 4], [4, 1], [1], [2], [3], [4]], Int[])
1∘2, 2∘3, 3∘4, 4∘1, 1, 2, 3, 4 -> 
julia> size_dict = uniformsize(eincode, 2)
Dict{Int64, Int64} with 4 entries:
  4 => 2
  2 => 2
  3 => 2
  1 => 2
julia> contraction_complexity(eincode, size_dict)
Time complexity: 2^4.0
Space complexity: 2^0.0
Read-write complexity: 2^4.643856189774725
\end{jllisting}
The \texttt{EinCode} is the default constructor for specifying the tensor network topology.
It takes indices for the input tensors as the first argument and output tensor as the second argument.
The \texttt{uniformsize} function is used to specify uniform dimension of each index, which returns a dictionary mapping each index to its dimension.
The \texttt{contraction\_complexity} function is used to compute the time complexity, space complexity, and read-write complexity of the contraction.
The time complexity is $2^4$ because the default contraction naively looks over all indices without creating intermediate tensors.
This leads to a large time complexity of $2^n$ for a $n$-index contraction.
Moreover, the non-binary contraction does not make use of BLAS libraries, leading to a large overhead.
In the following, we call the \texttt{optimize\_code} function to optimize the contraction order.
\begin{jllisting}
julia> optcode = optimize_code(eincode, size_dict, TreeSA())
SlicedEinsum{Int64, DynamicNestedEinsum{Int64}}(Int64[], 1, 1 -> 
├─ 4∘1, 4∘1 -> 1
│  ├─ 4∘1
│  └─ 2∘4, 1∘2 -> 4∘1
│     ├─ 3∘2, 3∘4 -> 2∘4
│     │  ├─ 2, 2∘3 -> 3∘2
│     │  │  ├─ 2
│     │  │  └─ 2∘3
│     │  └─ 4∘3, 4 -> 3∘4
│     │     ├─ 3∘4, 3 -> 4∘3
│     │     │  ⋮
│     │     │  
│     │     └─ 4
│     └─ 1∘2
└─ 1
)
julia> contraction_complexity(optcode, size_dict)
Time complexity: 2^5.087462841250339
Space complexity: 2^2.0
Read-write complexity: 2^6.108524456778169
\end{jllisting}
The first argument of the \texttt{optimize\_code} function is the \texttt{EinCode} to be optimized, the second argument is the size dictionary, and the third argument is the optimizer.
The \texttt{TreeSA} is a heuristic optimizer based on the local search method. It returns a \texttt{SlicedEinCode}, which is a contraction order with sliced indices.
Slicing is a technique to reduce the space complexity by looping over a subset of indices.
Here, since we did not set how many indices to be sliced, the first field of the \texttt{SlicedEinCode} is empty.
Since the demonstrated graph is too small, the time complexity is not reduced.
In the following subsection, we will introduce more methods for optimizing the contraction order.
They can be used to replace the \texttt{TreeSA()} method in the above example.

\subsection{Methods for optimizing contraction order}

\begin{figure}[ht]
    \centering
    \includegraphics[width=0.6\textwidth, trim={0 0 0 0}, clip]{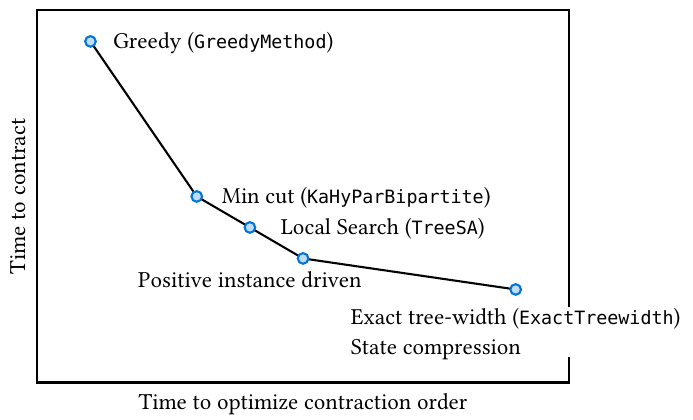}
    \caption{The time to contract the tensor network versus the time to optimize the contraction order. The method types implemented in \texttt{OMEinsum} are annotated in the parentheses.}
    \label{fig:comparison}
\end{figure}

\texttt{OMEinsum} includes various methods to automatically find a good contraction order and serves as a backend of \texttt{GenericTensorNetwork}.
We list some methods for optimizing contraction order in \Cref{fig:comparison}.
Two of the methods give the contraction order with the optimal space complexity, they are the exact tree-width solver and the state compression method that implemented in \texttt{TensorOperations.jl}~\cite{TensorOperations.jl}.
Both requires a time exponential in the number of tensors to optimize the contraction order, thus they are not suitable for large tensor networks with more than $50$ tensors.
For larger networks such as those from a quantum circuit~\cite{Markov2008}, probabilistic inference problem~\cite{martin2024probabilistic} or the combinatorial optimization~\cite{Liu2023, Liu2024},
we usually resort to faster heuristic methods.
There is a trade-off between the time for optimization and the quality of the contraction order.
The better contraction order is usually obtained at the cost of more time to optimize the contraction order.
In the following, we will introduce the methods implemented in \texttt{OMEinsum} in detail.

\subsubsection{Greedy method}

The Greedy method is one of the simplest and fastest method for optimizing the contraction order. 
The idea is to greedily select the pair of tensors with the smallest cost to contract at each step. 
In each step, for all possible pairs of tensors, the cost of the contraction is evaluated, which is defined as
\begin{equation}
    \mathcal{L}(T_i, T_j) = \text{size}(T_{i} * T_{j}) - \alpha (\text{size}(T_i) + \text{size}(T_j)),
\end{equation}
where $T_i$ and $T_j$ are the tensors to be contracted, $T_{i} * T_{j}$ is the intermediate tensor after contracting $T_i$ and $T_j$, and $\alpha$ is a parameter.
Then the pair with the smallest cost is selected and then contracted, which forms a new tensor.  
This process is repeated until all tensors are contracted. 
This method is fast, however it is easy to be trapped in local minima.

A variant of the greedy method is called the hyper-greedy method~\cite{Gray2021}, where in each step one samples according to the Boltzmann distribution given by $\mathcal{P}(T_i, T_j) = e^{-\mathcal{L}(T_i, T_j) / T}$ instead of directly selecting the pair with the smallest cost, where $T$ is the temperature.
If setting $T = 0$, the hyper-greedy method is equivalent to the greedy method.
In this case, it is possible for the process to escape from local minima.
Then the process is repeated multiple times and the best result is selected.

In \texttt{OMEinsum}, the greedy method is implemented as \texttt{GreedyMethod}, with three parameters: $\alpha$, $\texttt{temperature}$, and $\texttt{nrepeat}$.
Their default values are set to $0.0$, $0.0$, and $10$, respectively.
\begin{jllisting}
julia> using OMEinsum
julia> greedy = GreedyMethod(
    α = 0.0,                # the parameter of the cost function
    temperature = 0.0,      # the temperature of the hyper-greedy method
    nrepeat = 10,           # the number of trials
);
\end{jllisting}

\subsubsection{Local search method}

\begin{figure}[ht]
    \centering
    \includegraphics[width=0.6\textwidth, trim={0 0 0 0}, clip]{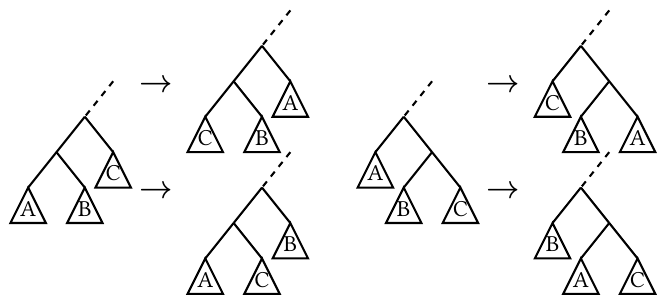}
    \caption{The four basic local transformations on the contraction tree, which preserve the result of the contraction.}
    \label{fig:tree_transform}
\end{figure}

The local search method~\cite{Kalachev2022} is a heuristic method based on the idea of simulated annealing.
The method starts from a random contraction order and then applies the following four possible transforms as shown in \Cref{fig:tree_transform}, which correspond to the different ways to contract three sub-networks:
\begin{equation*}
    \begin{split}
    &(A * B) * C = (A * C) * B = (C * B) * A, \\
    &A * (B * C) = B * (A * C) = C * (B * A),
    \end{split}
\end{equation*}
where we slightly abuse the notation ``$*$'' to denote the tensor contraction, and $A, B, C$ are the sub-networks to be contracted.
Due to the commutative property of the tensor contraction, such transformations do not change the result of the contraction.
Even through these transformations are simple, all possible contraction orders can be reached from any initial contraction order.
The local search method starts from a random contraction tree.
In each step, the above rules are randomly applied to transform the tree and then the cost of the new tree is evaluated, which is defined as
\begin{equation}
    \mathcal{L} = \texttt{tc} + w_s \texttt{sc} + w_{rw} \texttt{rwc},
\end{equation}
where $w_s$ and $w_{rw}$ are the weights of the space complexity and read-write complexity compared to the time complexity, respectively.
The the transfromation is accepted with a probability given by the Metropolis criterion, which is
\begin{equation}
    p_{\text{accept}} = \min\left(1, e^{-\beta \Delta \mathcal{L}}\right),
\end{equation}
where $\beta$ is the inverse temperature, and $\Delta \mathcal{L}$ is the difference of the cost of the new and old contraction trees.
During the process, the temperature is gradually decreased, and the process stop when the temperature is low enough.
Additionally, the \texttt{TreeSA} method supports the slicing technique.
When the space complexity is too large, one can loop over a subset of indices, and then contract the intermediate results in the end.
Such technique can reduce the space complexity, but slicing $n$ indices will increase the time complexity by $2^n$.

In \texttt{OMEinsum}, the local search method is implemented as \texttt{TreeSA} as shown in the following example.
\begin{jllisting}
julia> using OMEinsum
julia> treesa = TreeSA(
    sc_target = 20,          # the target space complexity
    βs = 0.01:0.05:15,       # the inverse temperatures
    ntrials = 10,            # the number of trials
    niters = 50,             # the number of iterations at each temperature
    sc_weight = 1.0,         # the relative weight of the space complexity
    rw_weight = 0.2,         # the relative weight of the read-write complexity
    initializer = :greedy,   # the initializer of the contraction order
    nslices = 0,             # the number of sliced indices
    fixed_slices = Any[],    # the manually fixed sliced indices
    greedy_config = GreedyMethod(0.0, 0.0, 1), # the method used as initializer
);
\end{jllisting}

\subsubsection{Binary partition}

Binary partition is a heuristic method~\cite{Gray2021} that builds the contraction tree in a top-down manner.
A tensor network can be represented as a weighted hypergraph, where the tensors are the vertices and the shared indices are the hyperedges.
The weight of the hyperedge associated with index $i$ is given by $w_i = \log_2 \text{size}(i)$, where $\text{size}(i)$ is the dimension of the index $i$.
Then by finding a balanced min cut on the hypergraph, we can partition the hypergraph into two parts, which represent two sub-networks to be contracted with each other.
The cutted edges are the indices to be contracted and the summation of the weights of the cutted edges is the cost of the contraction.
Thus, minimizing the total weight of the cutted edges is equivalent to minimizing the cost of the contraction, and by balancing the partition, the depth of the contraction tree is minimized.
Such bipartition is repeated recursively until the resulting sub-networks are small enough to be optimized by other simpler methods.

In the past few decades, the graph community has developed many algorithms for the balanced min cut problem and provided the corresponding software packages, such as \texttt{KaHyPar}~\cite{schlag2023high}.
In \texttt{OMEinsum}, the binary partition based methods are implemented as \texttt{KaHyParBipartite} and \texttt{SABipartite}, which solve the balanced min cut problem by \texttt{KaHyPar} and simulating annealing method, respectively.

\begin{jllisting}
julia> using OMEinsum, KaHyPar
julia> kahyparbiparatite = KaHyParBipartite(
    sc_target = 25,               # the target space complexity
    imbalances = 0.0:0.005:0.8,   # the imbalances of the partition
    max_group_size = 40,          # the maximum size of the partition
    sub_optimizer = GreedyMethod(0.0, 0.0, 10),  # the sub-optimizer
);
julia> sabiparatite = SABipartite(
    sc_target = 25,          # the target space complexity
    ntrials = 50,            # the number of trials
    βs = 0.1:0.2:14.9,       # the inverse temperatures
    niters = 1000,           # the number of iterations at each temperature
    max_group_size = 40,     # the maximum size of the partition
    sub_optimizer = GreedyMethod(0.0, 0.0, 10),   # the sub-optimizer
    initializer = :random,   # the initializer of the contraction order
);
\end{jllisting}

\subsubsection{Line graph tree decomposition}

Motivated by the results of Markov and Shi~\cite{Markov2008}, contraction order can be obtained by solving the tree decomposition of the line graph of the tensor network.
An index elimination order can be obtained from the tree decomposition, which gives a contraction order with contraction complexity upper bounded by the width of the tree decomposition~\cite{Gray2021, kalachev2021multi}.
Among all tree decompositions, the tree decomposition with minimal width is called the \textit{optimal tree decomposition}, its width is called the \textit{treewidth}.
The treewidth is an upper bound of time complexity of the single contraction step, thus the optimal tree decomposition is directly associated with the optimal contraction order.

\texttt{OMEinsum} provides exact treewidth based method \texttt{ExactTreewidth}, which implements the Bouchitté–Todinca algorithm~\cite{Bouchitté2001}.
Solving exact tree width for an arbitrary graph is NP-hard and takes exponential time, hence it only works for small networks.
Since the tree decomposition does not guarantee the contraction tree to be binary, a greedy method is used to convert the non-binary contractions into binary ones.

\begin{jllisting}
julia> using OMEinsum
julia> exacttreewidth = ExactTreewidth(
    greedy_config = GreedyMethod(nrepeat = 1),  # the sub-optimizer
);
\end{jllisting}

\section{Tensor networks contraction for solution space analysis}\label{sec:solution}
\subsection{Interface}
This section introduces how to use the generic tensor network for solution space analysis.
The generic tensor network method serves as a unified framework for solving constraint satisfaction problems through tensor networks, by linking the desired properties of the solution space with the algebraic operations employed in tensor network contraction.
Its Julia implementation is included in \texttt{GenericTensorNetworks.jl}, and the main feature is included in a single function \texttt{solve}:
\begin{lstlisting}
julia> using GenericTensorNetworks, Graphs#, CUDA
julia> solve(
           GenericTensorNetwork(          # convert the CSP problem to a tensor network
               IndependentSet(            # CSP problem: the independent set problem
                   square(20),            # 20x20 square lattice
                   UnitWeight(400)        # default: uniform weight 1
               );    
               optimizer = TreeSA(),      # contraction order optimizer
               openvertices = (),         # default: no open vertices
               fixedvertices = Dict()     # default: no fixed vertices
           ),
           PartitionFunction(0.0);        # wanted property: partition function at β = 0.0 (infinite temperature)
           usecuda=false                  # default: not using CUDA
       )
0-dimensional Array{Float64, 0}:
9.589790366629295e71
\end{lstlisting}

The \texttt{solve} function takes two positional arguments, one is the target CSP problem, the other is the wanted property. Here, the target problem is the unweighted independent set problem defined on a $20\times 20$ square lattice, and the wanted property is the partition function at infinite temperature $Z(\beta = 0)$. One extra keyword argument \texttt{usecuda=false} is used to specify not using CUDA for the tensor network contraction. What is happening behind the \texttt{solve} function is shown in \Cref{fig:gtn}, the CSP problem is converted to an energy model and then to a tensor network. Then the contraction order is optimized. Depending on the wanted property, the tensor network is contracted with the corresponding algebra. Here, the wanted property is the partition function, which corresponds to the standard real number algebra.
The full list of the properties and its associated algebra is shown in \Cref{tbl:generictypes}.
The output is an array.
Here, a $0$-dimensional array represents a scalar, which corresponds to the infinite temperature partition function, or the number of all independent sets.
From the output, we can see it is a extremely large number which is way beyond the data range of a 64-bit integer. 
Thus, here the output is represented as floating point numbers by default.

\begin{table}[t!]\centering
\begin{minipage}{\columnwidth}
   \small
\begin{tabular}{|m{12em} | m{15em} | m{21em}|}
   \hline
   \textbf{Property} & \textbf{Element type \& Fields}     & \textbf{Description} \\
   \hline
   \texttt{SizeMax()}    & \shortstack[l]{\\\texttt{Tropical} \\ \;\;- \texttt{n}: size}    & Largest solution size \\
   \hline
   \texttt{SizeMax(k)}    & {\shortstack[l]{\\ \texttt{ExtendedTropical\{k\}} \\ \;\;- \texttt{orders}: sizes}}    & Largest $k$ solution sizes \\
   \hline
   \texttt{PartitionFunction($\beta$)} & \texttt{Real}     & Partition function at inverse temperature $\beta$ \\
   \hline
   \texttt{CountingAll()} & \texttt{Real}     & Number of solutions of all sizes \\
   \hline
   \texttt{GraphPolynomial()} & \shortstack[l]{\\\texttt{Polynomial} \\ \;\;- \texttt{coeffs}: coefficients}     & Number of solutions at different sizes (positive) \\
   \hline
   \texttt{\shortstack[l]{GraphPolynomial(;\\ method=:laurent)}} & \shortstack[l]{\\\texttt{LaurentPolynomial} \\ \;\;- \texttt{coeffs}: coefficients\\ \;\;- \texttt{order}: lowest order}     & Number of solutions at different sizes \\
   \hline
   \texttt{CountingMax()} & \shortstack[l]{\\\texttt{CountingTropical} \\ \;\;- \texttt{n}: size\\ \;\;- \texttt{c}: counting}     & Number of solutions with largest size \\
   \hline
   \texttt{CountingMax(k)} & \shortstack[l]{\\\texttt{TruncatedPoly\{k\}} \\ \;\;- \texttt{coeffs}: coefficients\\ \;\;- \texttt{maxorder}: highest order}     & Number of solutions with largest $k$ sizes \\
   \hline
   \texttt{SingleConfigMax()} & \shortstack[l]{\\\texttt{\shortstack[l]{CountingTropical\{\\ \;\;\;Float64,\\ \;\;\;<:ConfigSampler\}}} \\ \;\;- \texttt{n}: size\\ \;\;- \texttt{c}: configuration\\ \;\;\;\;- \texttt{data}: vector}     & {One configuration for the largest solution size} \\
   \hline
   \texttt{SingleConfigMax(k)} & \shortstack[l]{\\\texttt{\shortstack[l]{ExtendedTropical\{k, \\\;\;\;<:CountingTropical\{\\\;\;\;\;\;\;Float64, \\\;\;\;\;\;\;<:ConfigEnumerator\}\}}} \\ \;\;- \texttt{orders}: sizes}     & {One configuration for each largest $k$ sizes} \\
   \hline
   \texttt{ConfigsMax()} & \shortstack[l]{\\\texttt{\shortstack[l]{CountingTropical\{\\\;\;\;Float64,\\\;\;\;<:ConfigEnumerator\}}} \\ \;\;- \texttt{n}: size\\ \;\;- \texttt{c}: configurations\\ \;\;\;\;- \texttt{data}: vector of vectors}     & {All solutions with largest size} \\
   \hline
   \texttt{ConfigsMax(k)} & \shortstack[l]{\\\texttt{\shortstack[l]{TruncatedPoly\{k\},\\\;\;\;<:ConfigEnumerator}} \\ \;\;- \texttt{orders}: sizes}     & {All solutions with largest $k$ sizes} \\
   \hline
   \texttt{ConfigsAll()} & \shortstack[l]{\\\texttt{ConfigEnumerator} \\ \;\;- \texttt{data}: vector of vectors}     & {All solutions} \\
   \hline
   \texttt{\shortstack[l]{\\ ConfigsAll(;\\ tree\_storage=true)}} & \shortstack[l]{\\\texttt{SumProductTree} \\ \;\;- \texttt{tag}: node type\\ \;\;- \texttt{count}: number of solutions\\ \;\;- \texttt{data}: vector \\\;\;- \texttt{left}: left child\\ \;\;- \texttt{right}: right child}     & {All solutions as an expression tree} \\
   \hline
\end{tabular}
\caption{Tensor element types and the independent set properties that can be computed using them.
Every property with \texttt{Max} in name has its \texttt{Min} counterpart. The \texttt{LaurentPolynomial} is used as return value if negative sizes are involved, e.g. in the case of spin glass. The \texttt{tree\_storage} option can be used in any property for configuration enumeration for memory saving.}
\label{tbl:generictypes}
\end{minipage}
\end{table}

\subsection{Tropical tensor network}

In the zero temperature limit, the logarithm of the partition function (\Cref{eq:partition}) can be rewritten as
\begin{equation}
    \lim_{\beta \to \infty} \log Z(G, \beta) = \max_{\mathbf{s}} \sum_{e \in E} h(e, s)\;,
\end{equation}
from which the tropical semiring algebra emerges~\cite{Liu2023}.
% The tropical semiring is a semiring with the addition operation defined as the maximum operation and the multiplication operation defined as the addition operation
The tropical semiring is a semiring of extended real numbers with the operations of minimum (or maximum) and addition replacing the usual operations of addition and multiplication, respectively.

By contracting the tensor network with the tropical semiring, we can get the maximum size.
The Julia programming language allows us to define a new type \texttt{Tropical} to represent the tropical semiring as demonstrated in the following code.

\begin{lstlisting}
julia> fieldnames(Tropical)  # fields of the Tropical number
(:n,)
julia> a, b = Tropical(2.0), Tropical(3.0)
(2.0ₜ, 3.0ₜ)
julia> a.n
2.0
julia> a + b, a * b         # `+` maps to `max`, `*` maps to `+`
(3.0ₜ, 5.0ₜ)
julia> zero(a), one(a)      # additive identity and multiplicative identity
(-Infₜ, 0.0ₜ)
julia> GenericTensorNetworks.generate_tensors(Tropical(1.0), IndependentSet(smallgraph(:petersen)))
25-element Vector{Array{Tropical{Float64}}}:
 [0.0ₜ 0.0ₜ; 0.0ₜ -Infₜ]
 [0.0ₜ 0.0ₜ; 0.0ₜ -Infₜ]
 ⋮
 [0.0ₜ, 1.0ₜ]
 [0.0ₜ, 1.0ₜ]
\end{lstlisting}

The property associated with the tropical semiring is \texttt{SizeMax}, as shown in \Cref{tbl:generictypes}. So we can use the following code to get the maximum size of independent sets of the Petersen graph.
\begin{lstlisting}
julia> net_petersen = GenericTensorNetwork(IndependentSet(petersen))
GenericTensorNetwork{IndependentSet{SimpleGraph{Int64}, Int64, UnitWeight}, OMEinsum.
DynamicNestedEinsum{Int64}, Int64}
- open vertices: Int64[]
- fixed vertices: Dict{Int64, Int64}()
- contraction time = 2^8.0, space = 2^4.0, read-write = 2^8.704
julia> res1 = solve(net_petersen, SizeMax())[]
4.0ₜ
\end{lstlisting}

\subsection{Solution space properties}
Continuing from the previous section, we introduce more properties that are associated with the counting and enumeration. The rigorous definition of the relevant algebra could be found in Ref.~\cite{Liu2024}.
The \texttt{SizeMax(k)} property is used for obtaining the largest $k$ sizes of independent sets of the Petersen graph with the extended tropical semiring. It is useful for obtaining the lowest lying solutions of the weighted problems.
\begin{lstlisting}
julia> res2 = solve(net_petersen, SizeMax(2))[]
ExtendedTropical{2, Tropical{Float64}}(Tropical{Float64}[4.0ₜ, 4.0ₜ])
\end{lstlisting}
The output is a vector of two same tropical numbers, which is due to the degeneracy of the largest two sizes. The \texttt{CountingMax} property is used for counting how many maximum size solutions there are.
\begin{lstlisting}
julia> res3 = solve(net_petersen, CountingMax())[]
(4.0, 5.0)ₜ
julia> res4 = solve(net_petersen, CountingMax(2))[]
30.0*x^3 + 5.0*x^4
\end{lstlisting}
When not specifying the number of solutions, the return type is \texttt{CountingTropical} and the counting is stored in the \texttt{c} field, otherwise the return type is \texttt{TruncatedPoly} and the counting is stored in the \texttt{coeffs} field.
The \texttt{CountingAll} property is used for counting all solutions.
\begin{lstlisting}
julia> res5 = solve(net_petersen, CountingAll())[]
76
\end{lstlisting}
This counting property is similar to the \texttt{PartitionFunction(0.0)} property, but uses integer with arbitrary precision as the return type. The \texttt{GraphPolynomial} property is used for counting the number of solutions at different sizes. Graph polynomial is an important concept in algebraic graph theory. Popular graph polynomials include the independence polynomial~\cite{Ferrin2014}, matching polynomial~\cite{Farrell1979}, and the chromatic polynomial~\cite{Read1968}. Generic tensor network method provides the \texttt{GraphPolynomial} property for CSP problems with integer sizes, which returns a polynomial that stores the number of solutions at different sizes in its coefficients.
\begin{lstlisting}
julia> res6 = solve(net_petersen, GraphPolynomial())[]
Polynomial(1 + 10*x + 30*x^2 + 30*x^3 + 5*x^4)
\end{lstlisting}
The \texttt{SingleConfigMax} property is used for obtaining one solution with the largest size. The return type is \texttt{CountingTropical} and the solution is stored in the \texttt{c} field.
In Julia, a type can be parameterized by another type, which is called a parameterized type. Here, \texttt{CountingTropical} is parameterized by \texttt{ConfigSampler}, which is a type that stores the solution as a binary string.
The \texttt{SingleConfigMax(k)} property is used for obtaining one solution for each largest $k$ sizes. The return type is \texttt{ExtendedTropical} and the solutions are stored in the \texttt{orders} field.
The solutions are represented as a vector of \texttt{CountingTropical} parameterized by \texttt{ConfigSampler}.
\begin{lstlisting}
julia> res7 = solve(net_petersen, SingleConfigMax())[]
(4.0, ConfigSampler{10, 1, 1}(1010000011))ₜ
julia> res8 = solve(net_petersen, SingleConfigMax(2))[]
ExtendedTropical{2, CountingTropical{Float64, ConfigSampler{10, 1, 1}}}(CountingTropical{Float64,
ConfigSampler{10, 1, 1}}[(4.0, ConfigSampler{10, 1, 1}(1001001100))ₜ, (4.0, ConfigSampler{10, 1, 1
}(0100100110))ₜ])
\end{lstlisting}
The \texttt{ConfigsMax} property is used for obtaining all solutions with the largest size. The return type is \texttt{CountingTropical} and the solutions are stored in the \texttt{c} field, which has type \texttt{ConfigEnumerator}.
The \texttt{ConfigsMax(k)} property is used for obtaining all solutions with the largest $k$ sizes. The return type is \texttt{TruncatedPoly} and the solutions are stored in the \texttt{coeffs} field.
Similarly, the \texttt{ConfigsAll} property is used for obtaining all solutions. The return type is an iterable of type \texttt{ConfigEnumerator}.
\begin{lstlisting}
julia> res9 = solve(net_petersen, ConfigsMax())[]
(4.0, {0010111000, 0101010001, 1010000011, 0100100110, 1001001100})ₜ
julia> res10 = solve(net_petersen, ConfigsMax(2))[]
{0100100010, 1001001000, 0100110000, 0000111000, 0101010000, 0001011000, 1010000010, 0010100010,
 1010001000, 0010110000, 0010011000, 0010101000, 1000000011, 0100000011, 1001000001, 0001010001,
 0100010001, 0101000001, 0010000011, 1010000001, 0010010001, 1000000110, 0000100110, 0100000110,
 1001000100, 1000001100, 0100100100, 0000101100, 0101000100, 0001001100}*x^3 + {0010111000,
 0101010001, 1010000011, 0100100110, 1001001100}*x^4
julia> res11 = solve(net_petersen, ConfigsAll())[]
{1000000010, 0100100010, 0000100010, 0100000010, 0000000010, 1001001000, 1001000000, 1000001000,
1000000000, 0100110000, 0000111000, 0000110000, 0101010000, 0001011000, 0001010000, 0100010000,
0000011000, 0000010000, 0100100000, 0000101000, 0000100000, 0101000000, 0001001000, 0001000000,
0100000000, 0000001000, 0000000000, 1010000010, 0010100010, 0010000010, 1010001000, 1010000000,
0010111000, 0010110000, 0010011000, 0010010000, 0010101000, 0010100000, 0010001000, 0010000000,
1000000011, 0100000011, 0000000011, 1001000001, 1000000001, 0101010001, 0001010001, 0100010001,
0000010001, 0101000001, 0001000001, 0100000001, 0000000001, 1010000011, 0010000011, 1010000001,
0010010001, 0010000001, 1000000110, 0100100110, 0000100110, 0100000110, 0000000110, 1001001100,
1001000100, 1000001100, 1000000100, 0100100100, 0000101100, 0000100100, 0101000100, 0001001100,
0001000100, 0100000100, 0000001100, 0000000100}
\end{lstlisting}
For larger solution spaces, the \texttt{tree\_storage} option can be used to save memory.
The return type is \texttt{SumProductTree} and the solutions are stored in the sum product expression tree so that the memory usage is significantly reduced.
To extract the solutions from the tree, we can use the \texttt{collect} function.
Interestingly, you can obtain a set of unbiased samples from the tree by using the \texttt{generate\_samples} function without enumerating the solutions.
\begin{lstlisting}
julia> res12 = solve(net_petersen, ConfigsAll(tree_storage=true))[]
+ (count = 76.0)
├─ + (count = 58.0)
│  ├─ + (count = 53.0)
│  ...
└─ * (count = 18.0)
julia> collect(res12)
76-element Vector{StaticBitVector{10, 1}}:
 0101000100
 0101000001
 ⋮
 0010110000
 0010100010
julia> generate_samples(res12, 3)
3-element Vector{StaticBitVector{10, 1}}:
 0101000100
 1000000110
 0010000010
\end{lstlisting}
This \texttt{tree\_storage} option is also available for the \texttt{ConfigsMax} and \texttt{ConfigsMin} properties.

\section{Applications}
\subsection{Hard square entropy constant}

The \textit{hard square entropy constant} is a quantity arises in statistical mechanics of hard-square lattice gases~\cite{Baxter1980, Pearce1988} to understand phase transitions for these systems.
It is defined as $\lim_{L\rightarrow \infty} F(L)^{1/L^2}$, where $F(L)$ is the number of independent sets of a given lattice dimensions $L \times L$.
Since the number of independent sets of a 1-dimensional lattice of length $n$ is the $n$th Fibonacci number, $F(L)$ form a well-known integer sequence (\href{https://oeis.org/A006506}{OEIS A006506}), which is thought as a two-dimensional generalization of the Fibonacci numbers.
Unlike the 1D Fibonacci numbers, $F(L)$ has no known polynomial time algorithm to compute.
The following code computes the hard square entropy constant for square lattice graphs of size $L \times L$.
\begin{lstlisting}
julia> F(L) = solve(GenericTensorNetwork(IndependentSet(square(L))), PartitionFunction(0.0))[]^(1/L^2)
\end{lstlisting}

Generic tensor network method allows us to compute the hard square entropy constant for square lattice graphs of size $L \times L$ with $L$ up to more than $39$.
\begin{figure}[h]
    \centering
    \includegraphics[width=0.5\textwidth, trim={0cm 0.5cm 0cm 0cm}, clip]{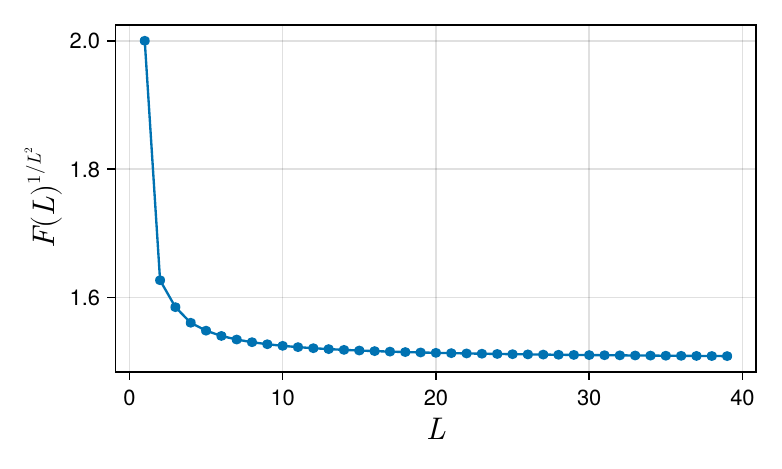}
    \caption{The entropy constant for the square lattice graph.}
    \label{fig:entropy}
\end{figure}

\subsection{Configuration enumeration and sampling}

The solution space analysis is crucial for designing better algorithms.
This section uses the independent set problem as an example to demonstrate the overlap gap property of the solution space, and how it can be different for different types of graphs.
How to characterize the hardness of a problem instance is one of the most important questions in the field of constraint satisfaction problems.

\begin{figure}[h]
    \centering
    \includegraphics[width=0.6\textwidth, trim={0cm 0cm 0cm 0cm}, clip]{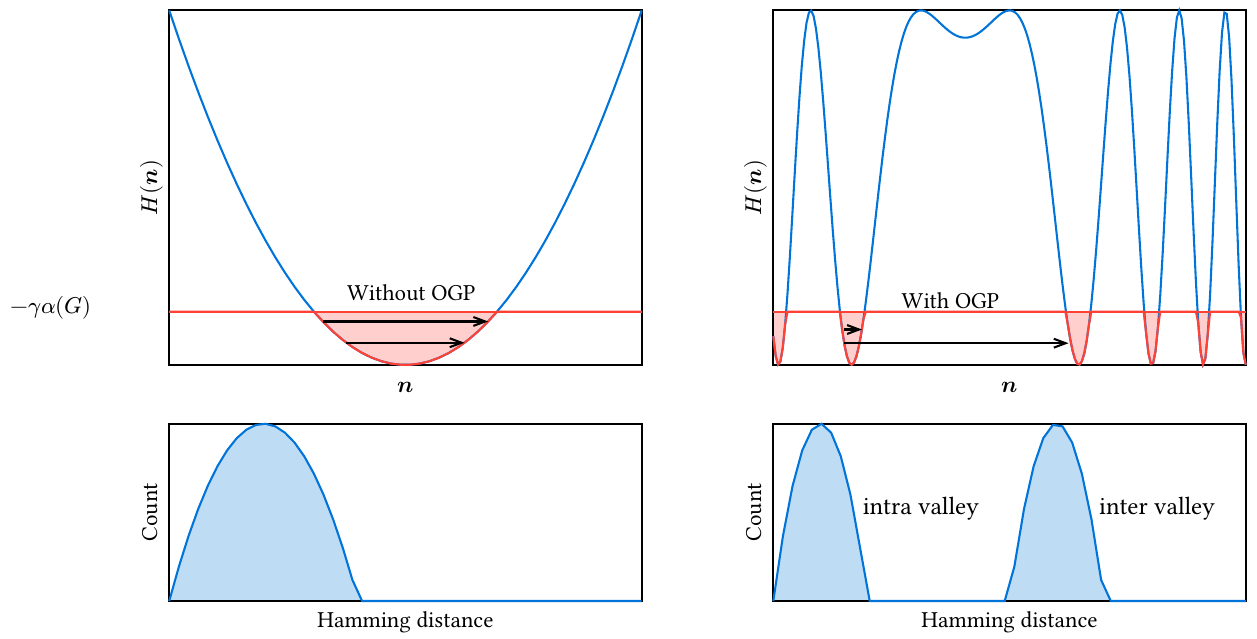}
    \caption{The pair-wise Hamming distance distribution of solutions at different energies for graphs with and without overlap gap property.}
    \label{fig:ogp}
\end{figure}

Unless the problem size is very small, enumeration of solutions at different energies is intractable.
To tell if solutions enumeration is feasible or not, we first count the number of solutions at different energies.
This task is closely related to the graph polynomials.
For example, the independence polynomial of a graph $G$ is a polynomial that counts the number of independent sets at different sizes
\begin{equation}
    I(G, x) = \sum_{k=0}^{|\alpha(G)|} c_k x^k
\end{equation}
where $c_k$ is the number of independent sets of size $k$ and $\alpha(G)$ is the size of the largest independent set of $G$.
If we can resolve the coefficients of the independence polynomial, we also know the number of independent sets at different sizes.
In the program, we use the property name \texttt{GraphPolynomial} to denote the graph polynomial, including the independence polynomial.

\begin{lstlisting}
julia> solve(GenericTensorNetwork(IndependentSet(petersen)), GraphPolynomial(; method=:finitefield))
0-dimensional Array{Polynomial{BigInt, :x}, 0}:
Polynomial(1 + 10*x + 30*x^2 + 30*x^3 + 5*x^4)
\end{lstlisting}
The \texttt{GraphPolynomial} property is accessible for problems with integer sizes.
It has a keyword argument \texttt{method}, which can be \texttt{:finitefield}, \texttt{:polynomial}, \texttt{:laurent}, \texttt{:fft} or \texttt{:fitting}. Here, we use the default method \texttt{:finitefield} to compute the independence polynomial of the Petersen graph, which has arbitrary precision. However, this method is unable to handle the problems with negative sizes, such as the spin glass. In such cases, we can use the \texttt{:laurent} method, which is based on the Laurent polynomial and can handle the negative polynomial orders.

\begin{figure}[h]
    \centering
    \includegraphics[width=0.6\textwidth, trim={0cm 0cm 0cm 0cm}, clip]{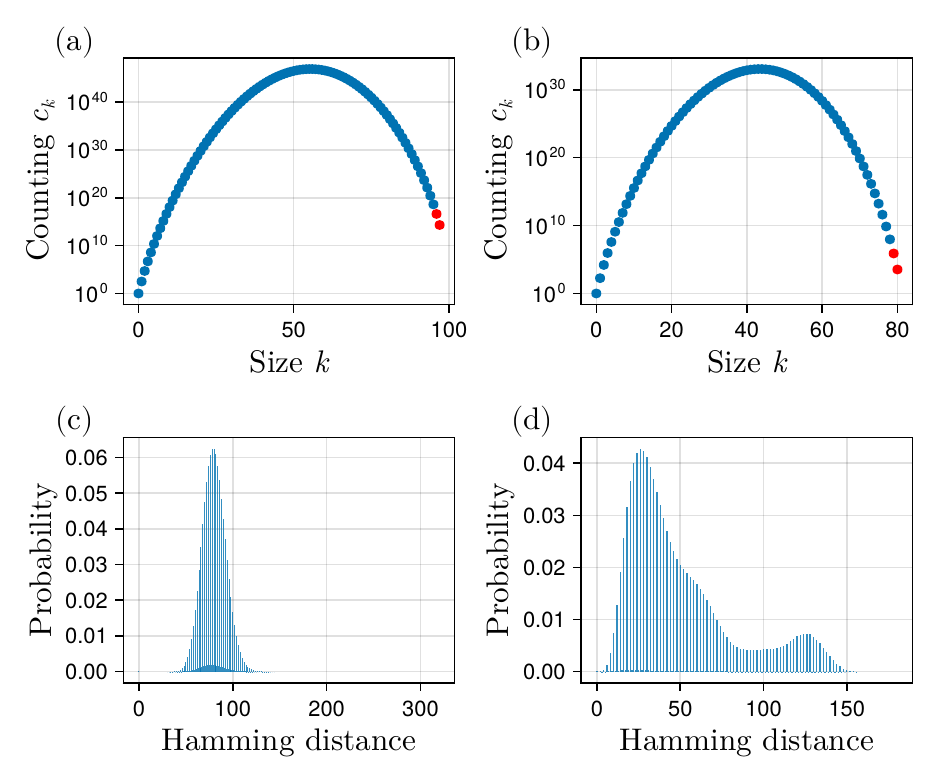}
    \caption{The number of solutions at different sizes for (a) a $20\times 20$ King's subgraph at filling $0.8$ and (b). a $180$ vertices $3$-regular graph. (c) and (d) are the Hamming distance distributions of the solutions with the largest sizes (red dots in (a) and (b)).}
    \label{fig:polynomial}
\end{figure}

We plot the number of solutions at different sizes for a $20\times 20$ King's subgraph at filling $0.8$ and a $180$ vertices $3$-regular graph in \Cref{fig:polynomial} (a) and (b).
The numbers of solutions at different sizes grow exponentially.
The solutions with the largest sizes are the most important, since they are highly related to the hardness to the problem.
However, even for the largest solutions, the number of solutions is still too large to be enumerated. We use the \texttt{tree\_storage} option of \texttt{ConfigsMax} property to save memory. The following code generates 10000 samples from the largest two solutions of a $20\times 20$ King's subgraph at filling $0.8$ and get the pair-wise Hamming distance distribution.
\begin{lstlisting}
julia> samples = generate_samples(sum(solve(
          GenericTensorNetwork(IndependentSet(ksg(20))),
          ConfigsMax(2; tree_storage=true)
        )[].coeffs), 10000);
julia> hamming = hamming_distribution(samples, samples);
\end{lstlisting}

Similar results can be obtained for the $180$ vertices $3$-regular graph. The results are shown in \Cref{fig:polynomial} (c) and (d). We can see that their Hamming distance distributions are very different. A random King's subgraph is more likely to have a small Hamming distance distribution, with a clear single peak. While the $3$-regular graph has a more uniform distribution, with a multiple peaks, which is an evidence of the overlap gap property.

\subsection{Reduce Factoring to Rydberg atoms for solution space analysis}

In this section, we introduce an extension of \texttt{ProblemReductions.jl}, the \href{https://github.com/QuEraComputing/UnitDiskMapping.jl}{\texttt{UnitDiskMapping.jl}} package, that contains extra reductions rules  allows users to reduce the factoring problem to the computational hard problems to that on a unit disk graph~\cite{Nguyen2023}.
The ultimate goal is to reduce a computational hard problem to the ground state finding problem of a physical system, such that computational hard problems can be solved by cooling a physical system to its ground state~\cite{Ebadi2022}.
Reducing the reduction overhead and understanding the change of solution space properties are crucial for designing better physics based algorithms.
To reduce the factoring problem to the weighted independent set problem on King's subgraph, the best known reduction requires $O(n^2)$ vertices~\cite{Nguyen2023}, where $n$ is the number of bits.
To use the extension, we just use two packages together. The following code reduces the factoring problem to the weighted independent set problem on King's subgraph and get the overlap gap property.

\begin{jllisting}
julia> using ProblemReductions, UnitDiskMapping, GenericTensorNetworks
julia> factoring_problem = Factoring(4, 5, 221);
julia> result = reduceto(IndependentSet{ProblemReductions.GridGraph{2}, Int, Vector{Int}}, factoring_problem);
julia> mapped_problem = target_problem(result)
IndependentSet{ProblemReductions.GridGraph{2}, Int64, Vector{Int64}}(ProblemReductions.GridGraph{2}([(14, 1), (16, 1), (32, 1), (34, 1), (50, 1), (52, 1), (18, 2), (36, 2), (54, 2), (6, 3)  …  (49, 56), (50, 56), (60, 56), (62, 56), (67, 56), (68, 56), (6, 58), (24, 58), (42, 58), (60, 58)], 2.8567113959936523), [2, 2, 2, 2, 2, 2, 2, 2, 2, 1  …  3, 3, 3, 3, 3, 3, 1, 1, 1, 1])
julia> problem_size(mapped_problem)
(num_vertices = 740, num_edges = 1559)
julia> config = findbest(mapped_problem, GTNSolver());
julia> result = ProblemReductions.read_solution(factoring_problem, extract_solution(result, first(config)))
(13, 17)
\end{jllisting}

One can easily verify that $13 \times 17 = 221$.

\begin{jllisting}
julia> samples = generate_samples(sum(solve(GenericTensorNetwork(mapped_problem), ConfigsMax(2; tree_storage=true))[].coeffs), 10000);
\end{jllisting}

\begin{figure}[h]
    \centering
    \includegraphics[width=0.6\textwidth, trim={0cm 0cm 0cm 0cm}, clip]{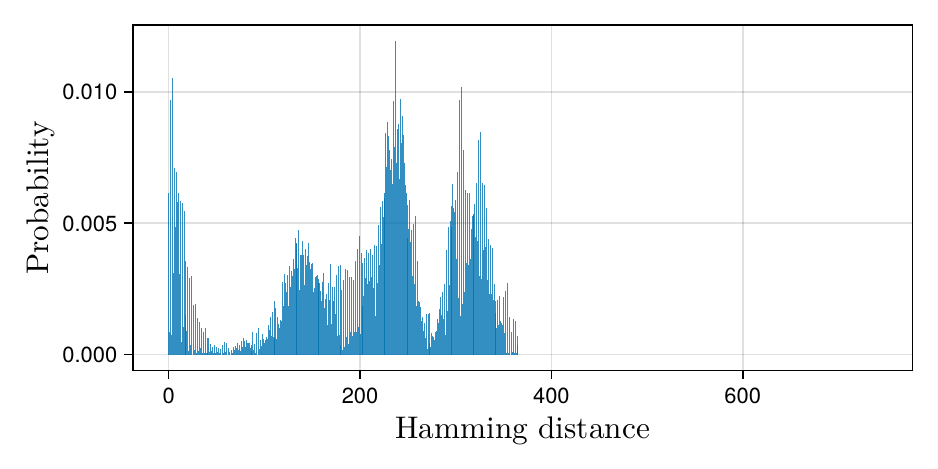}
    \caption{The Hamming distance distribution for the maximum 2 solutions of the independent set problem reduced from the factoring problem.}
    \label{fig:factoring}
\end{figure}

From the generated samples, we can visualize the Hamming distance distribution of the solutions with the largest 2 sizes in \Cref{fig:factoring}.
It turns out that the Hamming distance distribution has multiple peaks, which is an evidence of the overlap gap property.

\subsection{Ground state degeneracy of the Buckyball structure}

As a final example, we show how to use generic tensor network method to conquer the chanllenge problem released in the \href{https://github.com/QuantumBFS/SSSS}{Song Shan Lake Spring School 2019}. The problem statement is as follows.
% \linebreak\linebreak
% \fbox{\parbox{\textwidth}{
% In the Buckyball structure as shown in the figure, we attach an Ising spin  on each vertex. The neighboring spins interact with an anti-ferromagnetic coupling of unit strength.

% \centering{\includegraphics[width=0.2\textwidth]{images/fullerene.pdf}}

% \begin{enumerate}
%     \item Get $\ln(Z)/N$, where $N$ is the number of vertices, and $Z$ is the partition function at temperature $1.0$.
%     \item Count the ground state degeneracy.
% \end{enumerate}
% }
% }\linebreak\linebreak

\begin{problem}
    In the Buckyball structure (fullerene) illustrated in \Cref{fig:buckyball}, we assign an Ising spin to each vertex, with neighboring spins interacting through an anti-ferromagnetic coupling of unit strength. (a) Get $\ln(Z)/N$, where $N$ is the number of vertices, and $Z$ is the partition function at temperature $1.0$. (b) Count the ground state degeneracy.
    \begin{figure}[h]
        \centering
        \includegraphics[width=0.2\textwidth]{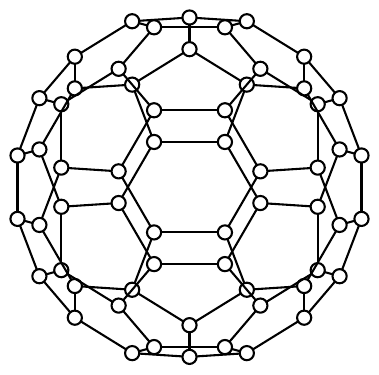}
        \caption{The Buckyball structure (fullerene).}
        \label{fig:buckyball}
    \end{figure}
\end{problem}

This problem can be easily solved by using the \texttt{GenericTensorNetworks.jl} package as follows.
\begin{lstlisting}
julia> using GenericTensorNetworks, Graphs, ProblemReductions
julia> function fullerene()  # construct the fullerene graph in 3D space
           th = (1+sqrt(5))/2
           res = NTuple{3,Float64}[]
           for (x, y, z) in ((0.0, 1.0, 3th), (1.0, 2 + th, 2th), (th, 2.0, 2th + 1.0))
               for (a, b, c) in ((x,y,z), (y,z,x), (z,x,y))
                   for loc in ((a,b,c), (a,b,-c), (a,-b,c), (a,-b,-c), (-a,b,c), (-a,b,-c), (-a,-b,c), (-a,-b,-c))
                       if loc not in res
                           push!(res, loc)
                       end
                   end
               end
           end
           return res
       end
fullerene (generic function with 1 method)
julia> fullerene_graph = UnitDiskGraph(fullerene(), sqrt(5)); # construct the unit disk graph
julia> spin_glass = SpinGlass(fullerene_graph, UnitWeight(ne(fullerene_graph)), zeros(Int, nv(fullerene_graph)));
julia> problem_size(spin_glass)
(num_vertices = 60, num_edges = 90)
julia> log(solve(spin_glass, PartitionFunction(1.0))[])/nv(fullerene_graph)
1.3073684577607942
julia> solve(spin_glass, CountingMin())[]
(-66.0, 16000.0)ₜ
\end{lstlisting}

The \texttt{UnitDiskGraph} type is a type that represents the unit disk graph, which is a graph that each vertex is a point in Euclidean space. Two vertices are connected by an edge if the distance between them is less than or equal to the given radius.
By solving the \texttt{PartitionFunction} and \texttt{CountingMin} properties, we get the logarithm of the partition function per vertex is $\approx 1.30737$ and the ground state degeneracy is $16000$, which is consistent with the expected results.

\section{Conclusion and future work}\label{sec:conclusion}
In this paper, we introduce the Julia ecosystem for solving constraint satisfaction problems with tensor networks, including the reduction between constraint satisfaction problems, the tensor network representation of constraint satisfaction problems, the tensor network contraction order optimizer, and the solution space analysis.
Source code is available in the following GitHub repositories.
\begin{itemize}
    \item \texttt{GenericTensorNetworks.jl}: \href{https://github.com/QuEraComputing/GenericTensorNetworks.jl}{https://github.com/QuEraComputing/GenericTensorNetworks.jl}
    \item \texttt{OMEinsum.jl}: \href{https://github.com/under-Peter/OMEinsum.jl}{https://github.com/under-Peter/OMEinsum.jl}
    \item \texttt{ProblemReductions.jl}: \href{https://github.com/GiggleLiu/ProblemReductions.jl}{https://github.com/GiggleLiu/ProblemReductions.jl}
\end{itemize}
Questions and contributions are welcome in the form of issues and pull requests.
In the future, we will continue to improve the performance of the tensor network contraction, and to extend the tensor network method to more constraint satisfaction problems through problem reduction.
At the same time, we have to warn the readers that the tensor network method is usually not the best choice when users only need a single solution, especially for a problem with a high dimension graph topology.
The branching algorithm works better as an exact solver in many cases~\cite{Fomin2013,Gao2024}. While for approximate solvers, local search and evolutionary algorithms~\cite{Lamm2017} can be more efficient.
The generic tensor network method is designed for counting the number of solutions and analyzing the solution space properties.

\section*{Acknowledgments}
We thank Chen-Guang Guan for actively contributing to the package \texttt{ProblemReductions.jl} in the open source promotion plan 2024.
We thank Andreas Peter and other contributors for their valuable contributions to \texttt{OMEinsum.jl} and \texttt{GenericTensorNetworks.jl} ecosystem.
This work is partially funded by the National Key R\&D Program of China (Grant No. 2024YFE0102500), National Natural Science Foundation of China (No. 12404568), the Guangzhou Municipal Science and Technology Project (No. 2023A03J00904), the Quantum Science Center of Guangdong-Hong Kong-Macao Greater Bay Area and the Undergraduate Research Project from HKUST(Guangzhou).

\bibliographystyle{plain}
\bibliography{ref}

% \appendix

% \section{Scripts to visualize the results}
% The following script is for visualizing multiple Hamming distance distributions in a grid.
% \begin{lstlisting}
% using CairoMakie, LinearAlgebra, GenericTensorNetworks

% function visualize_stats_grid(sample_grid)
%     fig = Figure()
%     for i in axes(sample_grid, 1), j in axes(sample_grid, 2)
%         samples = sample_grid[i, j]
%         dist = normalize(hamming_distribution(samples, samples), 1)
%         ax = Axis(fig[i, j])
%         @show dist
%         barplot!(ax, 0:length(dist)-1, dist)
%         ylims!(ax, 0, 0.2)
%     end
%     fig
% end

% # the Hamming distances of low enrgy configurations of random diagonal coupled graphs
% ksg_samples = [sample_low_energy_space(random_diagonal_coupled_graph(10, 10, 0.8)) for i=1:3, j=1:3]
% visualize_stats_grid(ksg_samples)

% # that for three regular graphs
% reg_samples = [sample_low_energy_space(random_regular_graph(100, 3)) for i=1:3, j=1:3]
% visualize_stats_grid(reg_samples)
% \end{lstlisting}

\end{document}